\documentclass[twocolumn,showpacs,preprintnumbers,amsmath,amssymb]{revtex4}

\usepackage{graphicx}
\usepackage{dcolumn}
\usepackage{bm}
\usepackage{color}

\begin{document}


\title{
Matter-wave dark solitons in a double-well potential}

\author{Ryoko Ichihara$^1$}
\author{Ippei Danshita$^2$}
\author{Tetsuro Nikuni$^2$}
\affiliation{
{$^1$Department of Physics, Waseda University, Shinjuku-ku, Tokyo 169-8555, Japan}
\\
{$^2$Department of Physics, Tokyo University of Science, Shinjuku-ku, Tokyo 162-8601, Japan}
}

\date{\today}

\begin{abstract}
We study stability of the first excited state of quasi-one-dimensional
Bose-Einstein condensates in a double-well potential, which is called ``$\pi$-state".
The density notch in the $\pi$-state can be regarded as a standing dark soliton.
From the excitation spectrum,
we determine the critical barrier height, above which the $\pi$-state is
dynamically unstable.
We find that the critical barrier height decreases monotonically as the number
of condensate atoms increases.
We also simulate the dynamics of the $\pi$-state by solving the time-dependent
Gross-Pitaevskii equation.
We show that due to the dynamical instability the dark soliton starts to move away from the trap center and exhibits a large-amplitude oscillation.
\end{abstract}

\maketitle
\section{Introduction}

Since the first discovery of Bose-Einstein condensates (BECs)
of alkali atoms in 1995~\cite{rf:anderson, rf:davis}, 
macroscopic quantum phenomena, 
which stem from the coherence of the many-body wavefunctions describing BECs, 
have been studied extensively. 
Systems of a BEC in a double-well potential are well suited for 
exploring various phenomena associated with the coherent nature, 
and vigorous studies in such systems in recent years have led 
to the observation of intriguing macroscopic quantum phenomena, 
such as the matter-wave interference~\cite{rf:andrews} and the 
Josephson effects~\cite{rf:albiez}.

The coherent nature of Bose-condensed atomic gases combined with 
their diluteness allows for the formulation of the BEC systems 
based on a non-linear Schr\"odinger equation, 
the so-called Gross-Pitaevskii (GP) equation~\cite{rf:BEC}. 
The static solutions of the GP equation describe a BEC not only
in the ground state but also in the macroscopically excited states, such as the stationary
quantized vortices~\cite{rf:fetter} and the standing dark solitons.
In particular, the dark solitons in BECs, 
which share fundamental properties with solitons in the nonlinear 
optics and the fluid mechanics, have attracted much 
attention~\cite{rf:burger, rf:denschlag, rf:becker}. 
Most recently, Becker {\it et al}.~observed experimentally 
dark solitons with long lifetime up to several seconds 
and opened up new possibility of detailed studies of the soliton 
dynamics~\cite{rf:becker}. 
Previous theoretical work~\cite{rf:muryshev, rf:feder} on 
stability of the dark soliton states 
in harmonic potentials with axial symmetry found that the dark soliton is stable when the
radial confinement is sufficiently strong, i.e. when the BEC is 
quasi-one-dimensional (1D). 
On the other hand, in 3D systems a dynamical instability, which is called 
the snake instability, causes the breakdown of the dark soliton, 
resulting in the formation of vortex rings~\cite{rf:denschlag, rf:feder}.

\begin{figure}[tb]
\begin{center}
\includegraphics[height=40mm]{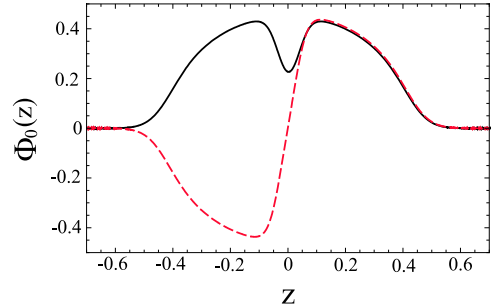}
\caption{\label{fig:pistate}(Color online)
The condensate wavefunctions for the ground state (solid line) and the $\pi$-state (dashed line) 
of BECs in a double-well potential.
We set the potential barrier $V_0/(\hbar\omega_z)=10.0$.
We consider condensates of ${{\rm ^{23}Na}}$ atoms, the mass 
is 23 amu (atomic mass units) and the atom number $N=2000$. 
The values of parameters are as follows: 
the harmonic-oscillator frequency is ${\omega_{z}/(2\pi)}=10$ Hz, 
the frequency of the radial confinement is ${\omega_{\perp}/(2\pi)}=250$ Hz
and the ${s}$-wave scattering length is ${a}_s=3$ nm.}
\end{center}
\end{figure}

The first excited state of a BEC in a double-well potential has 
a density notch and a phase kink at the center of the trap 
(see the dashed line in Fig.~\ref{fig:pistate}). 
This state is called $\pi$-state 
because the phase difference between the left and right condensates is $\pi$. 
In the sufficiently large condensate, 
the size of the density notch becomes of the order of the healing length. 
In this regime, the density notch can be regarded as a dark soliton.
Stability of the $\pi$-state of quasi-1D BECs has been previously 
studied only in two extreme limits; it is dynamically stable 
when the potential barrier is absent~\cite{rf:muryshev}, while it is dynamically 
unstable when the potential barrier height is much larger than the 
chemical potential, i.e. when the two-mode approximation is valid~\cite{rf:raghavan}. 
Although the transition from the dynamically stable $\pi$-state 
to the dynamically unstable $\pi$-state is expected to occur, 
no work to date has studied the intermediate transition region. 
In order to identify the transition, one needs linear stability 
analyses based on the Bogoliubov equations without 
using the two-mode approximation, which is not applicable in the transition region. 

In this paper, we study stability of the $\pi$-state at zero temperature in a wide 
range of the barrier height on the basis of the GP mean-field theory. 
First, we calculate the excitation energies of the $\pi$-state 
by solving the Bogoliubov equations numerically. 
The critical barrier height, above which the $\pi$-state 
is dynamically unstable, is determined from the condition of complex excitation energy. 
We will show that the critical barrier height decreases monotonically 
as the number of atoms increases. 
For comparison, we also calculate the lowest excitation-energy using 
the two-mode approximation. 
This calculation will confirm that the 
two-mode approximation completely fails to provide the critical 
barrier height except in the limit of small number of atoms. 
Secondly, we numerically simulate the dynamics of the $\pi$-state 
by solving the time-dependent GP equation. 
We will find that in the dynamically unstable region the condensate 
density notch corresponding to the dark soliton oscillates with 
a large amplitude around the center of the trap, while in the 
dynamically stable region the dark soliton remains stationary 
at the trap center during the time evolution.

The present paper is organized as follows.
In Sec.~\ref{sec:model}, we start with the model and formulation
based on the GP mean-field theory.
In Sec.~\ref{sec:stability}, we solve the time-independent GP equation
and the Bogoliubov equations numerically,
and we obtain the stability phase diagram.
In Sec.~\ref{sec:tdGP}, we numerically solve the time-dependent GP equation
to elucidate the decay process of the dynamically unstable condensates.
In Sec.~\ref{sec:summary}, we summarize our results.
 
\section{Model and Formulation}\label{sec:model}

We consider a BEC at zero temperature confined in a double-well potential.
The double-well potential is assumed to be composed of a harmonic trap 
with cylindrical symmetry and a Gaussian-shaped potential barrier,
\begin{eqnarray}
V_{\rm ext}({\bf r})=\frac{m\omega_{\perp}^2 \rho^2}{2}+\frac{m\omega_z^2 z^2}{2}
+V_0\,{\rm exp}\left(-\frac{z^2}{\sigma^2}\right),
\end{eqnarray}
where $\omega_{\perp}$ ($\omega_{z}$) is the frequency of the radial (axial) harmonic confinement,
$V_0$ is the barrier height and $\sigma$ is the width of the barrier.
In typical experiments, a potential barrier is created by a blue-detuned laser beam
and its height is controlled by the intensity of the laser beam.
We assume that $\hbar\omega_{\perp}$ is much larger than the chemical potential
so that we can justify the 1D treatment of this problem.
Accordingly, equilibrium and nonequilibrium properties of BECs
in a double-well potential are described by the 1D time-dependent GP equation for the condensate wavefunction $\Psi(z, t)$: 
\begin{eqnarray}
{\rm i}\hbar\frac{\partial}{\partial t}\Psi(z, t)
\!=\!
\left[\!
-\frac{\hbar^2}{2m}\frac{d^2}{dz^2}\!+\!V_{\rm ext}(z)\!+\!g|\Psi(z, t)|^2
\!\right]\!
\Psi(z, t),
\label{eq:tdGPE}
\end{eqnarray}
where
\begin{eqnarray}
V_{\rm ext}(z)=\frac{m\omega_z^2 z^2}{2}+V_0\,{\rm exp}\left(-\frac{z^2}{\sigma^2}\right).
\label{eq:external}
\end{eqnarray}
The coupling constant $g$ in 1D is given by $g=2\hbar^2 a_s/ma_{\perp}^2$,
where $a_s$ is the $s$-wave scattering length 
and $a_{\perp}\equiv (\hbar/m \omega_\perp)^{1/2}$ is the harmonic oscillator length of the radial confinement.

The static solution $\Psi_0(z, t)=\Phi_0(z) {\rm e}^{-{\rm i}\mu t/\hbar}$ of Eq. \!(\ref{eq:tdGPE}) satisfies the time-independent GP equation,
\begin{eqnarray}
\left[\!
-\frac{\hbar^2}{2m}\frac{d^2}{d z^2}-\mu+V_{\rm ext}(z)
+g|\Phi_0(z)|^2
\!\right]\!
\Phi_0(z) =0,\label{eq:sGPE}
\end{eqnarray}
where $\mu$ is the chemical potential.
$\Phi_0(z)$ satisfies the normalization condition,
\begin{eqnarray}
\int |\Phi_0(z)|^2 dz = N,
\end{eqnarray}
where $N$ is the number of condensate atoms.
Considering small oscillations around the static solution $\Phi_0(z)$, we write the condensate wavefunction as
\begin{eqnarray}
\Psi(z,t)
\!=\!
{\rm e}^{-{\rm i}\mu t/\hbar}
\!\left[
\Phi_0(z)+u(z){\rm e}^{-{\rm i}\varepsilon t/\hbar}
+v^{\ast}(z){\rm e}^{{\rm i}\varepsilon^{\ast} t/\hbar}
\right].
\label{eq:smallamp}
\end{eqnarray}
Substituting Eq. \!(\ref{eq:smallamp}) into Eq. \!(\ref{eq:tdGPE}) and linearizing it with respect to the quasiparticle amplitudes $u(z)$ and $v(z)$,
we obtain the Bogoliubov equations
\begin{eqnarray}
             \left(
                \begin{array}{cc}
                 h_0(z) & g[\Phi_0(z)]^2 \\
                 -g[\Phi_0^{\ast}(z)]^2 & -h_0(z)
                 \end{array}
               \right) 
               \left(
                 \begin{array}{cc}
                 u(z) \\ v(z)
                 \end{array}
               \right)
               = \varepsilon\left(
                 \begin{array}{cc}
                   u(z) \\ v(z)
                 \end{array}
               \right),  \label{eq:BdGE}
\end{eqnarray}
where we introduced the operator
\begin{eqnarray}
               h_0(z) = -\frac{\hbar^2}{2m}\frac{d^2}{d z^2}
               -\mu+V_{\rm ext}\left(z\right)+2g|\Phi_0(z)|^2.
\end{eqnarray}
The solutions of the Bogoliubov equations~(\ref{eq:BdGE}) determine the excitation energy $\varepsilon$.

We solve the time-independent GP equation~(\ref{eq:sGPE})
and the Bogoliubov equations~(\ref{eq:BdGE}) using the method introduced in Ref.~\cite{rf:edwards}.
We employ the harmonic oscillator eigenfunctions $\varphi_{\lambda}(z)$ as a basis set, where 
\begin{eqnarray}
\left(\!
-\frac{\hbar^2}{2m}\frac{d^2}{d z^2}+\frac{m \omega_z^2}{2}z^2
\!\right)\!
\varphi_{\lambda}(z) = \varepsilon_{\lambda} \varphi_{\lambda}(z),\\
\varepsilon_{\lambda}=\left(\lambda +\frac{1}{2}\right)\hbar\omega_z
\quad(\lambda=0, 1, 2\ldots).
\end{eqnarray}
They satisfy the orthonormality relation
\begin{eqnarray}
\int dz \varphi^{\ast}_{\lambda}(z)\varphi_{\lambda'}(z) = \delta_{\lambda \lambda'}.   \label{eq:orthonormality}
\end{eqnarray}
We expand the static solution as
\begin{eqnarray}
\Phi_0(z)=\sqrt{N} \sum_{\lambda}a_{\lambda}\varphi_{\lambda}(z), 
\label{eq:hermite}
\end{eqnarray}
and insert Eq.~(\ref{eq:hermite}) into Eq.~(\ref{eq:sGPE}).
Multiplying Eq.~(\ref{eq:sGPE}) by $\varphi_{\lambda}^{\ast}(z)$ 
and making use of Eq.~(\ref{eq:orthonormality}), 
we obtain
\begin{eqnarray}
(\varepsilon_{\lambda}-\mu)a_{\lambda}+V_{0}\sum_{\lambda'}\left[\int dz 
\exp\biggl(-\frac{z^2}{\sigma^2}\biggr)a_{\lambda'}\varphi^{\ast}_{\lambda}\varphi_{\lambda'}
\right]
\nonumber\\
+gN\sum_{\lambda_{1},\lambda_{2},\lambda_{3}}\left[\int dz \varphi^{\ast}_{\lambda}\varphi^{\ast}_{\lambda_{1}}\varphi_{\lambda_{2}}\varphi_{\lambda_{3}}\right]a^{\ast}_{\lambda_{1}}a_{\lambda_{2}}a_{\lambda_{3}}=0. \label{eq:newtonraphson}
\end{eqnarray}
One can solve these nonlinear simultaneous equations for $a_{\lambda}$ 
by the Newton-Raphson method.
Since the condensate wavefunction of the $\pi$-state is an odd function, we take
only the harmonic oscillator eigenfunctions with odd parity in order to obtain 
the $\pi$-state.
For instance, in Fig.~\ref{fig:pistate} we show
the ground state and the $\pi$-state for $N=2000$ and $V_0/(\hbar\omega_z)=10.0$. 
Subsequently, by substituting the condensate wavefunction $\Phi_0(z)$ into 
the Bogoliubov equations~(\ref{eq:BdGE}) and expanding $u$ and $v$ in terms of the harmonic oscillator eigenfuntions as 
$u(z)=\sum_{\lambda}u_{\lambda}\varphi_{\lambda}(z)$ and 
$v(z)=\sum_{\lambda}$$v_{\lambda}\varphi_{\lambda}(z)$, 
we obtain the simultaneous linear equations for $u_{\lambda}$ and $v_{\lambda}$.
One can obtain the excitation energy $\varepsilon$ by diagonalizing the coefficient matrix of the linear equations.

In the following sections, we study stability of the $\pi$-state 
in a double-well potential.
As shown in Fig.~\ref{fig:pistate}, 
the major difference between the ground state and the $\pi$-state
is that the $\pi$-state has a notch with
zero density and the phase jump at the center of the trap.
The notch of the condensate density
with the phase jump of $\pi$ is regarded as a dark soliton,
when the number of atoms is large (Thomas-Fermi limit~\cite{rf:BEC}).
In the case of the ground-state BEC, the relation between the excitation energies of the ground state and the
barrier height in a double-well potential has been studied~\cite{rf:salasnich, rf:danshita}.
It was found that the lowest excitation-energy exhibits the crossover
from the dipole mode to the Josephson plasma mode as the barrier height 
increases~\cite{rf:danshita}. 

\section{Stability Analysis}\label{sec:stability}
\begin{figure}[tb]
\begin{center}
\includegraphics[height=50mm]{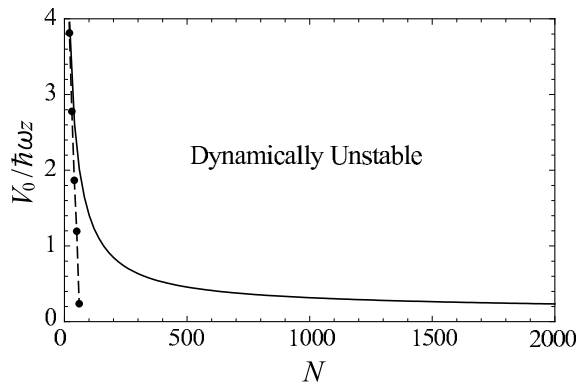}
\caption{\label{fig:SPD}
Stability phase diagram calculated from the Bogoliubov equations~(\ref{eq:BdGE}) (solid line) and the two-mode approximation (dashed line).}
\end{center}
\end{figure}
\begin{figure}[tb]
\begin{center}
\includegraphics[height=60mm]{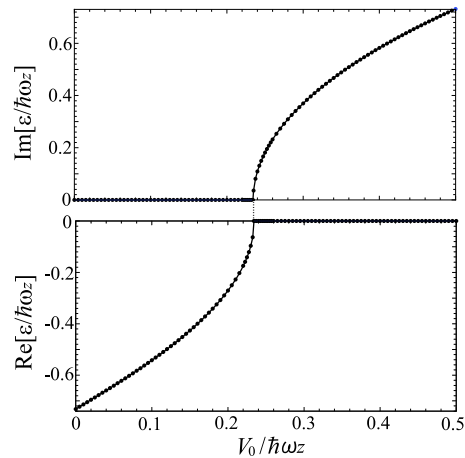}
\caption{\label{fig:N2000-1}
The lowest or imaginary eigenvalue of the Bogoliubov equations~(\ref{eq:BdGE})
as a function of $V_0/(\hbar\omega_z)$ with the fixed atom number $N=2000$.}
\end{center}
\end{figure}
\begin{figure}[tb]
\begin{center}
\includegraphics[height=50mm]{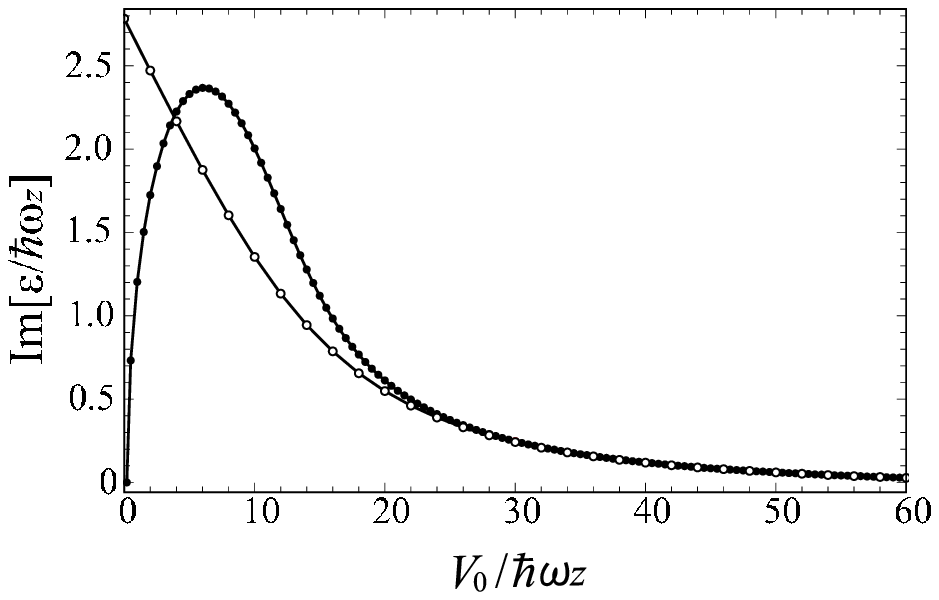}
\caption{\label{fig:N2000-2}
The closed circles represent the excitation energy for $N=2000$ as a function 
of $V_0/(\hbar\omega_z)$.
$\varepsilon_{\rm tma}$ is also plotted by the opened circles.}
\end{center}
\end{figure}
\begin{figure}[tb]
\begin{center}
\includegraphics[height=60mm]{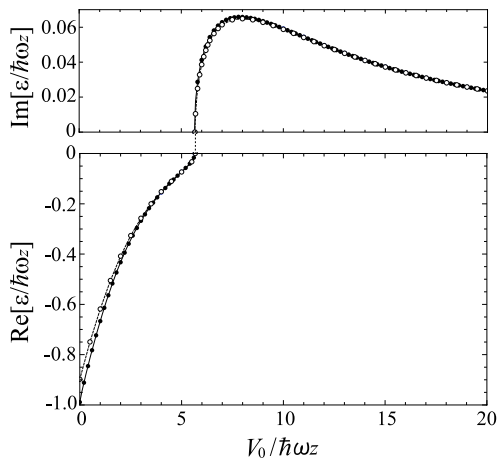}
\caption{\label{fig:N10}
The closed circles represent the excitation energy for $N=10$ as a function 
of $V_0/(\hbar\omega_z)$.
$\varepsilon_{\rm tma}$ is also plotted by the opened circles.}
\end{center}
\end{figure}
In this section, we solve the time-independent GP equation~(\ref{eq:sGPE}) 
and the Bogoliubov equations~(\ref{eq:BdGE}) numerically and obtain the stability phase diagram.
The stability of BECs can be studied by calculating the excitation energy.
The appearance of excitations with complex energies signals the dynamical instability whereby the amplitude of the condensate fluctuation grows exponentially in time.
On the other hand, the appearance of excitations with negative energies signals the Landau (energetic) instability, which means that the solution $\Phi_0$ of the time-independent GP equation does not correspond to a local minimum in the energy landscape.
Nevertheless, the Landau instability cannot destabilize the system in the limit of zero temperature because of the absence of dissipative processes~\cite{rf:muryshev,rf:konabe,rf:iigaya}.
In other words, the system is dynamically stable under the condition ${\rm Re}[\varepsilon]<0$ and ${\rm Im}[\varepsilon]=0$.

Fig.~\ref{fig:SPD} shows the stability phase diagram in the $(N, V_{0})$-plane.
The solid line represents the critical barrier height 
above which ${\rm Im}[\varepsilon] \neq 0$ so that the system is dynamically unstable.
One sees that the critical barrier height monotonically decreases as $N$ increases.

From the numerical solution of Eq.~(\ref{eq:BdGE}), we find that there is only one eigenstate with a negative excitation energy in the dynamically stable region (as also found in a special case of $V_{0}=0$ in Ref.~\cite{rf:muryshev}). 
This excitation corresponds to the oscillation of the density notch around
the trap center.
The reason for the negative energy for this excitation can be easily understood for the case of $V_0=0$.
In this case, it is clear that the potential energy for the density notch is the highest at the center of the trap.
Thus, the deviation of the density notch from the potential maximum
lowers the energy, resulting in the negative excitation energy.
As $V_{0}$ increases for a fixed value of $N$, the absolute value of 
this negative excitation energy decreases because the potential energy
for the density notch also decreases.
As $V_{0}$ increases from the critical point, the imaginary part 
of the excitation energy develops and takes the maximum value at a certain point. 
With increasing barrier height $V_{0}$ further, the imaginary part of 
the excitation energy decreases; it approaches zero asymptotically 
in the limit $V_{0}\rightarrow \infty$. 
In this limit, the two condensates separated by the potential barrier 
are completely independent of each other and thus the system becomes
dynamically stable again.

Here we discuss the case of $N=2000$, where the mean-field interaction
is so strong that the density notch of the condensate at the center
of the trap can be regarded as a dark soliton.
The excitation energy of the mode relevant to the stability 
of the $\pi$-state is shown in Figs.~\ref{fig:N2000-1} and ~\ref{fig:N2000-2}.
The trends mentioned above are clearly seen in these figures.
The critical barrier height is ${V_{0}}/{(\hbar\omega_{z})}\sim 0.24$, 
while the barrier height that gives the maximum value of 
Im$[\varepsilon]$ is ${V_{0}}/{(\hbar\omega_{z})}\sim$6.0.

We next calculate the excitation energy 
within the two-mode approximation.
This approximation is well known to provide analytical insights into BECs 
in a double-well potential when the barrier height is sufficiently 
large compared to the chemical potential~\cite{rf:raghavan,rf:danshita,rf:ananikian}.
The analytical expression of the excitation energy can be written as
\begin{eqnarray}
\varepsilon_{\rm{tma}}=\sqrt{2K(2K-UN)},
\end{eqnarray}
which is easily derived from the Josephson Hamiltonian~\cite{rf:raghavan}.
The coupling energy ${K}$ and the capacitive energy ${U}$ are defined as
\begin{eqnarray}
K=\frac{E_{\rm{an}}-E_{\rm{sm}}}{2},\quad\quad
U=2\frac{d\mu_{\rm{sm}}}{dN},
\end{eqnarray}
where ${E_{\rm{sm(an)}}}$ is the energy of the condensate of 
the symmetric (antisymmetric) state, and ${\mu_{\rm{sm}}}$ is 
the chemical potential of the symmetric state.
The energy of the condensate is given by
\begin{eqnarray}
E&=&\int_{-\infty}^{\infty} dz \Phi_0^{\ast}(z)\left[-\frac{\hbar^2}{2m} \frac{d^2}{dz^2}
+V_{\rm ext}(z)+\frac{g}{2}|\Phi_0(z)|^2\right]\nonumber\\
&&\times \Phi_0(z).
\end{eqnarray}

The dashed line in Fig.~\ref{fig:SPD} represents the transition point 
between the dynamically stable and unstable regions 
determined from the condition ${\rm Im}[\varepsilon_{\rm{tma}}] \neq 0$.
In Fig.~\ref{fig:N10}, $\varepsilon_{\rm tma}$ gives almost the same value 
as the excitation energy $\varepsilon$ determined from the Bogoliubov equations~(\ref{eq:BdGE}) even near
the critical barrier height, since the transition point is located in the region
where the barrier height is larger than the chemical potential.
In contrast, it is clearly seen in Fig.~\ref{fig:N2000-2} that 
$\varepsilon_{\rm tma}$ for $N$=2000 completely disagrees with $\varepsilon$ 
in the vicinity of the transition point, where $V_0\ll \mu$.
Thus, the two-mode approximation fails to provide a correct phase boundary
in the large $N$ region.

\section{Dynamics of the Condensate Wave Function}\label{sec:tdGP}
In this section, we solve the time-dependent GP equation~(\ref{eq:tdGPE})
numerically to elucidate the decay process
of the dynamically unstable condensates.
We use the split operator fast-Fourier transform method. 
As in Fig.~\ref{fig:N2000-1}, we set the number of atoms as $N=2000$.
In order to trigger the instability, we need to add 
a small random noise to the initial state as 
\begin{eqnarray}
\Psi(z, 0) \equiv \Phi_0(z)+\delta \Psi(z).
\end{eqnarray}
The amplitude of the random noise is chosen to be 
$\int dz |\delta \Psi|^2 \sim N \times 10^{-4}$.
In actual experiments, there always exists such a small noise due to thermal and
quantum fluctuations of the condensates or the vibration of the external field potential. 
\begin{figure}[tbhp]
\begin{center}
\includegraphics[height=85mm]{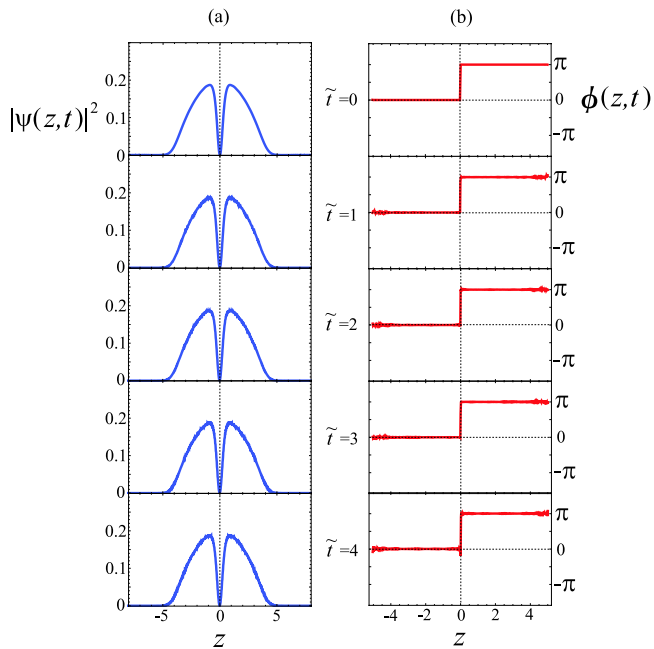}
\caption{\label{fig:TDGPv0}(Color online)
Time evolution of (a)~the density and (b)~the phase of the condensate wavefunction obtained by solving the time-dependent GP equation starting with the $\pi$-state. The barrier height is $V_{0}/(\hbar\omega_z)$=0, where the system is dynamically stable.
The dimensionless time is defined by $\tilde{t}\equiv t\omega_z$.}
\end{center}
\end{figure}
\begin{figure}[tbhp]
\begin{center}
\includegraphics[height=85mm]{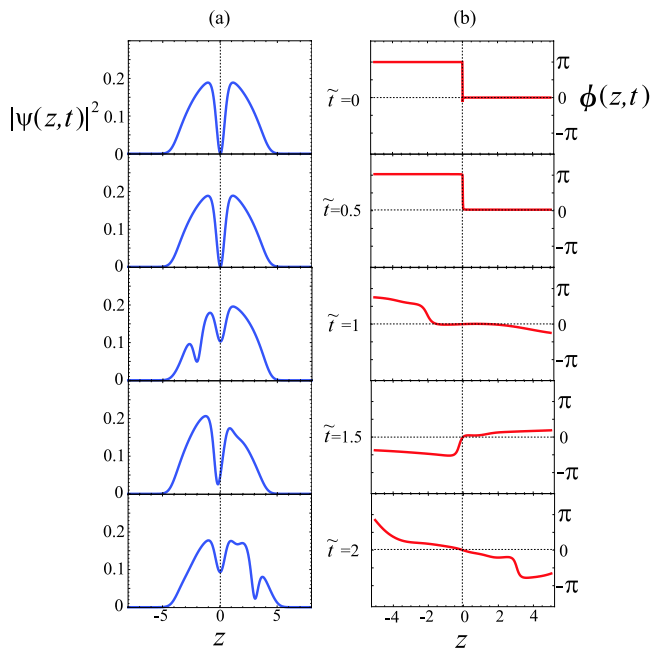}
\caption{\label{fig:TDGPv6}(Color online)
Time evolution of (a)~the density and (b)~the phase of the condensate wavefunction in the $\pi$-state obtained by solving the time-dependent GP equation in the dynamically unstable region $V_{0}/(\hbar\omega_z)$=6.0.
The dimensionless time is defined by $\tilde{t}\equiv t\omega_z$.}
\end{center}
\end{figure}
\begin{figure}[tbhp]
\begin{center}
\includegraphics[height=50mm]{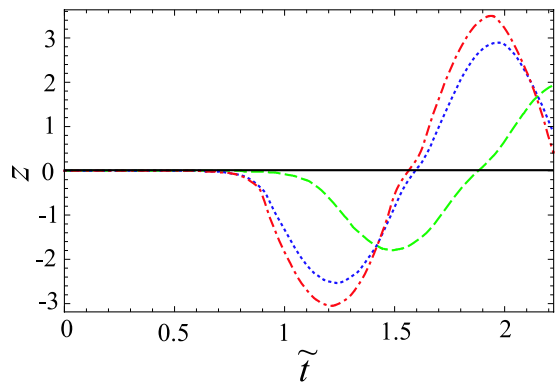}
\caption{\label{fig:TDGPkink}(Color online)
The kink position $z$ of the gray soliton as a function of the dimensionless time $\tilde{t}\equiv t\omega_z$. 
The cases of $V_0/(\hbar\omega_z)$=0, 0.1 and 0.2 (in the dynamically stable region) 
correspond to the solid line. 
The dashed, dotted and dashed-dotted lines represent the kink positions for 
$V_0/(\hbar\omega_z)$=2.0, $V_0/(\hbar\omega_z)$=4.0 and $V_0/(\hbar\omega_z)$=6.0 
(in the dynamically unstable region), respectively.}
\end{center}
\end{figure}

In Figs.~\ref{fig:TDGPv0} and ~\ref{fig:TDGPv6},
we plot the time evolution of the density $|\Psi(z,t)|^2$ and phase $\phi(z,t)$ of the condensate.
The phase $\phi(z,t)$ is an important quantity in understanding the soliton dynamics 
since the phase jump at the density notch 
$\Delta\phi$ is related to the velocity of the gray soliton $v$ by
\begin{eqnarray}
\Delta\phi=2\arccos\frac{v}{c},
\label{eq:PJ}
\end{eqnarray}
where $c$ is the sound velocity~\cite{rf:BEC}.
Moreover, we show the position of the dark soliton 
as a function of time in Fig.~\ref{fig:TDGPkink}.

As seen in Fig.~\ref{fig:TDGPv0} and the solid line in Fig.~\ref{fig:TDGPkink}, when
$V_0/(\hbar\omega_z) = 0$, the density notch remains stationary at the trap center
during the time evolution and the phase jump remains $\pi$.
Thus, the system is dynamically stable.

When the barrier height is above the critical value as shown in Fig.~\ref{fig:TDGPv6} ($V_0/(\hbar\omega_z) = 6.0$) and the dashed-dotted line in Fig.~\ref{fig:TDGPkink}, the density notch starts 
to move away from the center of the trap after a certain time 
$t\sim ({\rm Im}[\varepsilon/\hbar])^{-1}$
owing to the dynamical instability.
The density notch then develops a large-amplitude oscillation.
One can see that the notch moves with larger speed as the absolute value 
of the phase jump becomes smaller, consistent with Eq.~(\ref{eq:PJ}).

Since the condensate dynamics in the dynamically unstable region
is distinctively different from that in the stable region as discussed above,
one expects that the transition to the dynamically unstable region 
can be identified experimentally.
We suggest the following procedure to observe the transition in experiments:
at first one could engineer the $\pi$-state in a double-well trap by means of
the phase imprinting techniques~\cite{rf:burger,rf:carr}.
Subsequently, one could tune the barrier height by controlling the
intensity of the blue-detuned laser beam.
When the barrier height is so small that the system is in the stable region,
one should observe a dark soliton localized at a trap center.
Setting the barrier height to be large enough to exceed the critical value,
one should observe a large-amplitude oscillation of a dark soliton. 
Thus, the transition should be identified by measuring the time evolution
of the condensate density profile.

\section{Conclusion}\label{sec:summary}

In conclusion, we have studied stability of the first excited state 
of Bose-Einstein condensates in a double-well potential on the basis of 
the Gross-Pitaevskii mean-field theory.
From the excitation energies obtained from the Bogoliubov equations, 
we determined the critical barrier height above which the system is dynamically unstable.
The stability phase diagram was obtained against
 the barrier height $V_0$ and the number of condensate atoms $N$. 
 We found that the critical barrier height monotonically decreases as $N$ increases.

Solving the time-dependent GP equation numerically, 
we also studied the time evolution of the condensate in the $\pi$-state. 
Our simulation results show that the condensate density notch 
in the dynamically unstable region develops a large-amplitude oscillation; 
this behavior is significantly different from that in the dynamically 
stable region.

\begin{acknowledgements}
The authors would like to thank D. Yamamoto, S. Tsuchiya, and S. Kurihara for fruitful comments and discussions.
T. Inoue's assistance in numerical calculations is greatly acknowledged.
I. D. acknowledges support from a Grant-in-Aid from JSPS.
\end{acknowledgements}

\end{document}